\documentclass[a4paper,11pt]{article}
\usepackage{pos,enumitem,float,pifont}
\usepackage{xcolor, soul}

\usepackage{tikz}
\usetikzlibrary{arrows.meta}
\usetikzlibrary{decorations.markings}
\tikzstyle midarrow=[postaction={decorate,decoration={markings,
      mark=at position 0.53 with {\arrow{#1}},}}]

\graphicspath{{Figures/}}
\title{Using Gradient Flow to Renormalise Matrix Elements for Meson Mixing and Lifetimes}
\ShortTitle{Gradient Flow Renormalisation for Mixing and Lifetimes}

\author*[a]{Matthew Black}
\author[b]{Robert Harlander}
\author[c,d,e,f]{Fabian Lange}
\author[g]{Antonio Rago}
\author[b]{\\ Andrea Shindler}
\author[a]{Oliver Witzel}

\affiliation[a]{
    Center for Particle Physics Siegen, Theoretische Physik 1,
    Naturwissenschaftlich-Technische Fakult\"at,
    Universit\"at Siegen, 57068 Siegen, Germany
  }
\affiliation[b]{
    Institute for Theoretical Particle Physics and Cosmology,
    RWTH Aachen University, 52056 Aachen, Germany
  }
\affiliation[c]{
    Physik-Institut, 
    Universität Zürich, 
    Winterthurerstrasse 190, 8057 Zürich, Switzerland
  }
\affiliation[d]{
    Paul Scherrer Institut, 
    5232 Villigen PSI, Switzerland
  }
\affiliation[e]{
    Institut f\"ur Theoretische Teilchenphysik,
    Karlsruhe Institute of Technology,
    Wolfgang-Gaede-Stra{\ss}e 1, 76128 Karlsruhe, Germany
  }
\affiliation[f]{
    Institut f\"ur Astroteilchenphysik,
    Karlsruhe Institute of Technology,
    Hermann-von-Helmholtz-Platz 1, 76344 Eggenstein-Leopoldshafen, Germany
  }
\affiliation[g]{
    IMADA and Quantum Theory Center,
    University of Southern Denmark, Odense, Denmark
  }

\emailAdd{matthew.black@uni-siegen.de}

\abstract{
Neutral meson mixing and meson lifetimes are theory-side parametrised in terms four-quark operators which can be determined by calculating weak decay matrix elements using lattice Quantum Chromodynamics.
While calculations of meson mixing matrix elements are standard, determinations of lifetimes typically suffer from complications in renormalisation procedures because dimension-6 four-quark operators can mix with operators of lower mass dimension and, moreover, quark-line disconnected diagrams contribute.

We present work detailing the idea to use fermionic gradient flow to non-perturbatively renormalise matrix elements describing meson mixing or lifetimes, and combining it with a perturbative calculation to match to the $\overline{\rm MS}$ scheme using the \sftx.
}

\FullConference{%
  The 40th International Symposium on Lattice Field Theory (Lattice 2023)\\
  July 31st - August 4th, 2023\\
  Fermi National Accelerator Laboratory\\[8pt]
  {\rm Preprints: SI--HEP--2023--23, P3H-23-076, ZU-TH 68/23, PSI-PR-23-38, TTP23-049, TTK-23-27}}

\def\gev{\,\text{Ge\hspace{-0.1em}V}}
\def\mev{\,\text{Me\hspace{-0.1em}V}}
\def\fm{\,\text{fm}}

\def\MSb{\overline{\rm MS}}
\def\sftx{{short-flow-time expansion}}

\begin{document}
\maketitle

\section{Introduction}
The phenomenology of $B$ physics has become a rich and diverse study area at both dedicated $B$ factories such as Belle and BABAR and also more general collider experiments such as those at the LHC \cite{HFLAV:2022wzx}.
Over many years, there has been enormous efforts put in by the experimental community to increase the precision of measurements of the decays and properties of $B$ mesons (see e.g.~Ref.~\cite{Lenz:2022rbq}), and thus to fully leverage this success the precision of theoretical predictions for these decays and properties should similarly increase.
In particular, the behaviour of neutral meson mixing provides a key insight into CP violation in the Standard Model (SM) and can help constrain elements of the Cabibbo-Kobayashi-Maskawa (CKM) quark mixing matrix which further tests the SM and aids in searches for new physics. 
Furthermore, the lifetime of a particle is one of its fundamental properties and thus is of great importance in testing the underlying theory for consistency.
On the theory side, the lifetimes of $B$ mesons are determined in the framework of the heavy quark expansion (HQE) where the result is described in terms of a series expansion in $1/m_b$ of perturbative Quantum Chromodynamics (QCD) contributions and non-perturbative $\Delta B=0$ matrix elements; for a review, see e.g.~Ref.~\cite{Lenz:2014jha}.
In fact, multiple calculations of the $\Delta B=2$ matrix elements have already been carried out on the lattice \cite{Carrasco:2013zta,Aoki:2014nga,Gamiz:2009ku,Bazavov:2016nty,Dowdall:2019bea}~(and also using QCD sum rules \cite{Grozin:2016uqy,Grozin:2017uto,Grozin:2018wtg,Kirk:2017juj,King:2019lal}).
For recent overviews, see \cite{Tsang:2022smt,Chobanova:2022yll}; in addition RBC/UKQCD and JLQCD presented preliminary results for on-going work using an RI-SMOM renormalisation scheme \cite{Boyle:2021kqn,Erben:2023talk,Tsang:2023talk}. 
When considering the ratio of $B_s^0$ over $B^0$ mixing, the SU(3) breaking parameter $\xi$ can be defined \cite{Bernard:1998dg}. Since renormalisation factors and other uncertainties cancel, the lattice determination of $\xi$ is typically more precise \cite{Carrasco:2013zta,Aoki:2014nga,Gamiz:2009ku,Bazavov:2016nty,Dowdall:2019bea,Boyle:2018knm} and $\xi$ is frequently used in global CKM unitarity triangle fits \cite{CKMfitter,UTfit:2006vpt,UTfit:2022hsi}.

The history of the $\Delta B=0$ four-quark matrix elements is less thorough.
After early quenched studies \cite{DiPierro:1998ty,DiPierro:1999tb}~and preliminary unquenched results \cite{Becirevic:2001fy}~in the early 2000s, the topic received little attention from the lattice community until recently with some interest in lifetimes ratios and baryonic decays \cite{Lin:2022fun}.
In the meantime, there have been predictions using QCD sum rules \cite{King:2021jsq,Kirk:2017juj}.
These matrix elements present additional challenges for a lattice calculation due to contributions from disconnected diagrams where the signal-to-noise ratio worsens. 
Moreover, mixing with operators of lower mass dimension occurs in standard renormalisation procedures.

In the following we outline a non-perturbative renormalisation scheme utilising the gradient flow \cite{Narayanan:2006rf,Luscher:2009eq,Luscher:2010iy} and the \sftx~\cite{Luscher:2011bx,Suzuki:2013gza,Makino:2014taa,Monahan:2015lha} in which operator mixing is absent. 
A perturbative matrix is required to match each quantity to the $\MSb$ scheme, likely circumventing some of the difficulty in calculating $\Delta B=0$ four-quark matrix elements. 
The method is first tested for $\Delta F=2$ matrix elements where results can be verified against the literature. 
Our approach is similar to work by Suzuki et al.~applying the \sftx~to neutral Kaon mixing or the determination of the energy-momentum tensor \cite{Taniguchi:2020mgg,Suzuki:2021tlr,Suzuki:2020zue}.

\section{Gradient Flow and Short-Flow-Time Expansion}
The gradient flow \cite{Narayanan:2006rf,Luscher:2009eq,Luscher:2010iy,Luscher:2011bx,Luscher:2013cpa,Luscher:2013vga,Luscher:2014kea,DelDebbio:2013zaa,Shindler:2022tlx}~has become a well-known tool in lattice simulations with common usage for e.g.~scale setting. 
One introduces an auxiliary dimension, the flow time $\tau$ [$\gev^{-2}$] which acts as a UV regulator and provides a well-defined smearing of gauge and fermion fields through the first-order differential equations
\begin{align}
    \partial_\tau B_\mu(\tau,x) &= {\cal D}_\nu(\tau)G_{\nu\mu}(\tau,x), \quad B_\mu(0,x) = A_\mu(x), \\
    \partial_\tau\chi(\tau,x) &= {\cal D}^2(\tau)\chi(\tau,x), \quad~~~~~~\, \chi(0,x) = q(x),
\end{align}
where $G_{\nu\mu}(\tau)=\partial_\nu B_\mu(\tau)-\partial_\mu B_\nu(\tau)+[B_\nu(\tau),B_\mu(\tau)]$ is the flowed gluon field strength tensor, ${\cal D}_\nu(\tau)=\partial_\nu+[B_\nu(\tau),\cdot]$ is the flowed covariant derivative, $A_\mu,q$ are the regular gauge and fermion fields respectively and $B_\mu(\tau),\chi(\tau)$ are those extended in the flow time.
Operators evolved along positive gradient flow time are removed of UV divergences and are renormalised within a gradient flow (GF) scheme.

An effective Hamiltonian expressed normally as a sum of operators ${\cal O}_m$ and their Wilson coefficients $C_m$ can be rewritten in terms of these `flowed' operators $\tilde{\cal O}_n(\tau)$ and similarly `flowed' Wilson coefficients $\tilde{C}_n(\tau)$:
\begin{equation}
    {\cal H}_{\rm eff} = \sum_m C_m{\cal O}_m = \sum_n\tilde{C}_n(\tau)\tilde{\cal O}_n(\tau),
\end{equation}    
where the flow-time dependence of the operators cancels with that of the coefficients \cite{Monahan:2015lha,Harlander:2020duo,Harlander:2022tgk,Suzuki:2020zue}.
In the \sftx, one can relate the `flowed' operators to the regular ones as
\begin{equation}
    \tilde{\cal O}_n(\tau) = \sum_m\zeta_{nm}(\tau){\cal O}_m + O(\tau) \implies \sum_n\zeta_{nm}^{-1}(\mu,\tau)\langle\tilde{\cal O}^{\rm GF}_n\rangle(\tau) = \langle{\cal O}^{\MSb}_m\rangle(\mu),
\end{equation}
where higher-dimensional operators are accompanied by higher powers of $\tau$ and are expected to be negligible for small $\tau$ \cite{Luscher:2011bx,Suzuki:2013gza,Makino:2014taa,Monahan:2015lha}.
The perturbatively-calculated matrix $\zeta_{nm}^{-1}(\mu,\tau)$ matches the GF renormalised operators to the $\MSb$ scheme.

For a general heavy quark field $F$, we focus for now only on the bag parameter of the $\Delta F=2$ four-quark operator 
\begin{equation}
    {\cal O}_1 = (\bar F\gamma_\mu(1-\gamma_5)q)(\bar F\gamma_\mu(1-\gamma_5)q).
\end{equation}
This is well-studied in the literature and is the only contributor to $\Delta M$ in the SM.
In the future, we will extend our study to consider the full SUSY basis of $\Delta F=2$ four-quark dimension-six operators as well as $\Delta F=0$.

The bag parameters are defined as ratios of the three-point matrix element of a four-quark operator ${\cal O}_i$ to its vacuum insertion approximation.
For a pseudoscalar meson state $P$ with mass $m$ and decay constant $f$, the bag parameter of ${\cal O}_1$ is defined, at leading order, as
\begin{equation}
    B_1 = \frac{\langle P|{\cal O}_1|P\rangle}{\frac83 m^2f^2}.
    \label{eq:B1}
\end{equation}
The perturbative calculations used in this work are described in \cite{Harlander:2022tgk,Harlander:future}.
At next-to-next-to-leading order (NNLO), the perturbative matching from GF to $\MSb$ schemes for the bag parameter $B_1$ with number of flavours $n_f$ is given by
\begin{equation}
    \begin{aligned}
        \zeta_{B_1}^{-1}(\mu,\tau) = 1 &+ \frac{a_s}{4}\left(-\frac{11}{3} - 2L_{\mu \tau}\right) \\
                                              &+ \frac{a_s^2}{43200}\bigg[-2376 - 79650L_{\mu \tau} - 24300L_{\mu \tau}^2 + 8250n_f + 6000\,n_f\,L_{\mu \tau} \\ 
                                                                                                                             &\qquad\qquad~+ 1800\,n_f\,L_{\mu \tau}^2 - 2775\pi^2 + 300\,n_f\,\pi^2 - 241800\log2 \\
                                                                                                                             &\qquad\qquad~+ 202500\log3 - 110700\,{\rm Li}_2\left(\frac14\right)\bigg],
    \end{aligned}
\end{equation}
where $L_{\mu\tau}=\log(2\mu^2\tau)+\gamma_E$ and $a_s=\alpha_s/\pi$.
The final result for $B_1$ in the $\MSb$ scheme is given by
\begin{equation}
    B_1^{\MSb}(\mu) = \lim_{\tau\to0}\zeta_{B_1}^{-1}(\mu,\tau)B_1^{\rm GF}(\tau).
\end{equation}

\section{Lattice calculation}
\vspace{-10pt}
We will consider six RBC/UKQCD $2{+}1$-flavour domain-wall fermion (DWF) and Iwasaki gauge field ensembles with three lattice spacings $a\sim 0.11$, $0.08$, $0.07\fm$ (determined by RBC/ UKQCD~\cite{Blum:2014tka,Boyle:2017jwu,Boyle:2018knm}) and pion masses $\in[267,433)\mev$. 
Light and strange quarks are simulated with the Shamir DWF action \cite{Kaplan:1992bt,Shamir:1993zy,Furman:1994ky,Blum:1996jf} with $M_5=1.8$. 
These ensembles are listed in Table~\ref{tab:ensembles}. 
\begin{table}[th]
\[
  \begin{array}{ccccccccccccc}
  & L & T & L_s &  a^{-1}\!/\!\gev & am_l^\text{sea} & am_s^\text{sea} 
  & am_s^{\rm val} & M_\pi/\!\mev & \text{\# cfgs} & \text{\# sources}\\\hline
\text{C1} & 24 & 64 & 16 & 1.785 & 0.005 & 0.040 & 0.03224 & 340 & 101 & 32\\
\text{C2} & 24 & 64 & 16 & 1.785 & 0.010 & 0.040 & 0.03224 & 433 & 101 & 32\\[1.2ex]
\text{M1} & 32 & 64 & 16 & 2.383 & 0.004 & 0.030 & 0.02477 & 302 &  79  & 32\\
\text{M2} & 32 & 64 & 16 & 2.383 & 0.006 & 0.030 & 0.02477 & 362 &  89  & 32\\
\text{M3} & 32 & 64 & 16 & 2.383 & 0.008 & 0.030 & 0.02477 & 411 &  68  & 32\\[1.2ex]
\text{F1S}& 48 & 96 & 12 & 2.785 & 0.002144 & 0.02144 & 0.02167 & 267 & 98 & 24
  \end{array}
  \]
  \caption{RBC/UKQCD ensembles used in the discussed simulations~\cite{Allton:2008pn,Aoki:2010dy,Blum:2014tka,Boyle:2017jwu}.
    $am_l^\text{sea}$ and $am_s^\text{sea}$ are the light and strange sea quark masses and $M_\pi$ is the unitary pion mass. 
    $am_s^{\rm val}$ are the valence strange quark masses, set to the physical mass.}
  \label{tab:ensembles}
\end{table}

Heavy quarks are simulated using stout-smeared gauge fields \cite{Morningstar:2003gk}~and the M\"obius DWF action \cite{Brower:2012vk}, where the mass has been tuned to the physical charm on each ensemble through the $D_s$ pseudoscalar meson \cite{ParticleDataGroup:2022pth}.
Using a similar setup as Ref.~\cite{Boyle:2018knm}, all propagators are generated with Z2-noise wall sources where the number of sources and smearing parameters are listed in Table~\ref{tab:ensembles}; Gaussian smearing is also applied for the strange quarks.

In the following, we use exploratory results obtained on the C1, C2, and M1 ensembles. 
While testing the validity of our method, we remove the additional complications of extrapolations in the valence sector, studying only strange and charm quarks at their physical values.
As such, we currently consider the short-distance contributions to `neutral $D_s$' meson mixing.

\begin{figure}[th]
    \centering
    \includegraphics[width=0.6\textwidth]{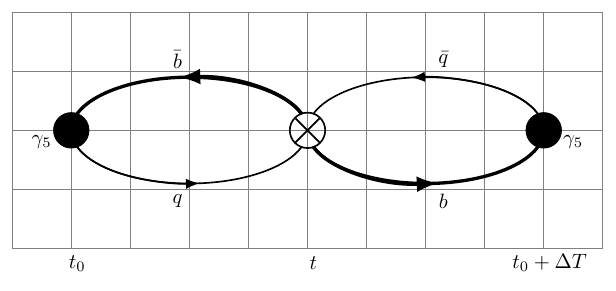}
    \caption{\label{fig:3ptfn} Quark line diagram of the lattice setup calculating the three-point correlation function with $\Delta F=2$ four-quark operator insertion at time $t$ between two sources placed at $t_0$ and $t_0+\Delta T$.}
\end{figure}

On the lattice, this is obtained in the large $t$ and $\Delta T$ limit by the ratio of correlation functions,
\begin{equation}
    R_1(t,\Delta T,\tau) = \frac{C_{{\cal O}_1}^{\rm 3pt}(t,\Delta T,\tau)}{\frac83 C_{AP}^{\rm 2pt}(t,\tau)C_{PA}^{\rm 2pt}(\Delta T-t,\tau)} \to B_1^{\rm GF}(\tau),
\end{equation}
where $t$ is the Euclidean time and $\Delta T$ is the separation of the two sources used in the three-point function as shown in Figure~\ref{fig:3ptfn}, and $C_{AP}^{\rm 2pt}(t,\tau),C_{PA}^{\rm 2pt}(\Delta T-t,\tau)$ are the two-point functions with the pseudoscalar current at the sources and the flowed axial current at the sink.
In this pilot study we only consider $\Delta T=28$ for all data analysed so far.
Dependence on the flow time $\tau$ is written here explicitly as the above ratio is evaluated for propagators taken at discrete steps in the flow.
The Runge-Kutta evolution of the gradient flow is performed with $\epsilon=0.01$.
For small flow, measurements of two- and three-point functions are taken at steps of $\varepsilon=0.1$ in lattice units, with this `coarsening' to $\varepsilon=0.4$ for $\tau/a^2>5$.
  
\section{First Results}
\begin{figure}[th]
    \centering
    \includegraphics[width=0.48\textwidth]{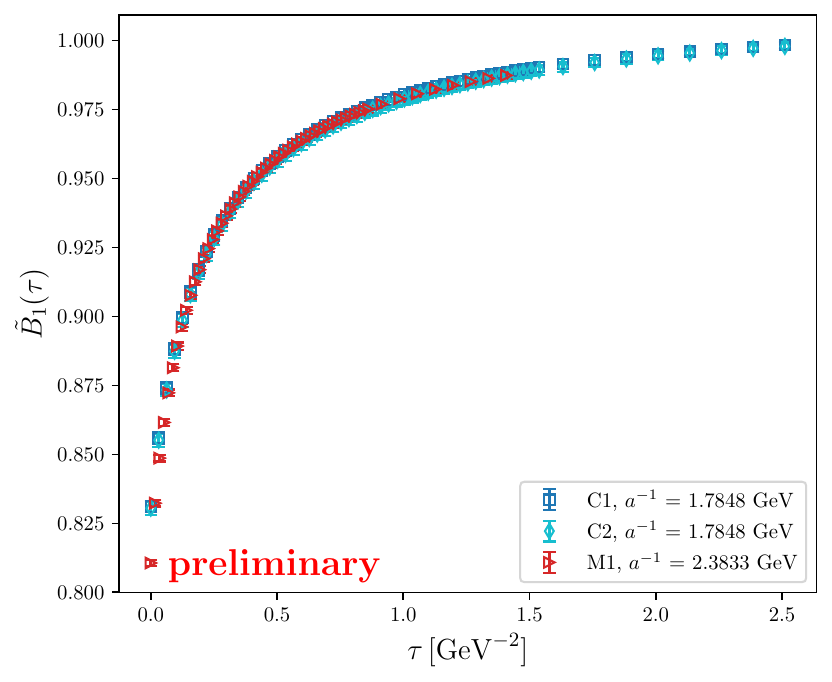}
    \includegraphics[width=0.48\textwidth]{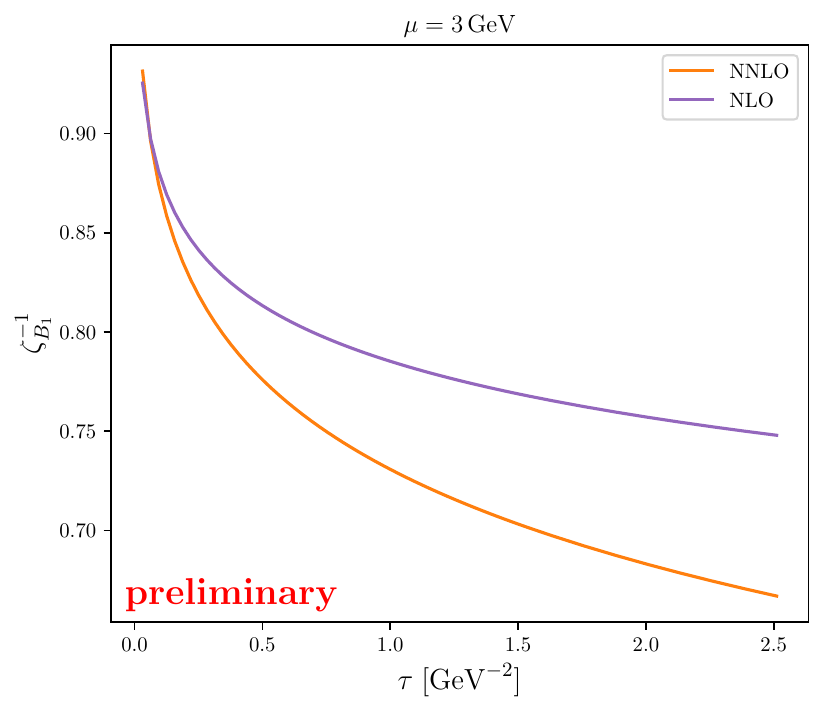}
    \caption{\label{fig:B1_GF} Evolution of the $\Delta F=2$ bag parameter $B_1^{\rm GF}(\tau)$ along the gradient flow for ensembles C1, C2, M1 in physical units with statistical errors only (left); perturbative GF-to-$\MSb$ matching coefficient $\zeta^{-1}_{B_1}(\mu,\tau)$ as a function of flow time $\tau$ with $\mu=3\,$GeV at both NLO and NNLO from Refs.~\cite{Harlander:2022tgk,Harlander:future} (right).}
\end{figure}

In Figure~\ref{fig:B1_GF}, we present our first results for the $\Delta F=2$ bag parameter $B_1^{\rm GF}(\tau)$ and its GF-to-$\MSb$ matching coefficient at both NLO and NNLO, expressed with the renormalisation scale $\mu=3\,$GeV both as functions of the gradient flow time $\tau$ in physical units.
The plot on the left shows the dependence of the lattice data on the GF time $\tau$ converted to physical units. The clear overlap of the data from different ensembles indicates a mild continuum limit for large enough flow times ($\tau \gtrsim 0.2\,{\rm GeV}^{-2}$) where the results become GF renormalised; for smaller flow times, the continuum limit may however carry a substantial systematic uncertainty. The plot on the right shows the perturbative matching with a clear difference between the next-to-leading order (NLO) and NNLO results. 

\begin{figure}[th]
    \centering
    \includegraphics[width=0.48\textwidth]{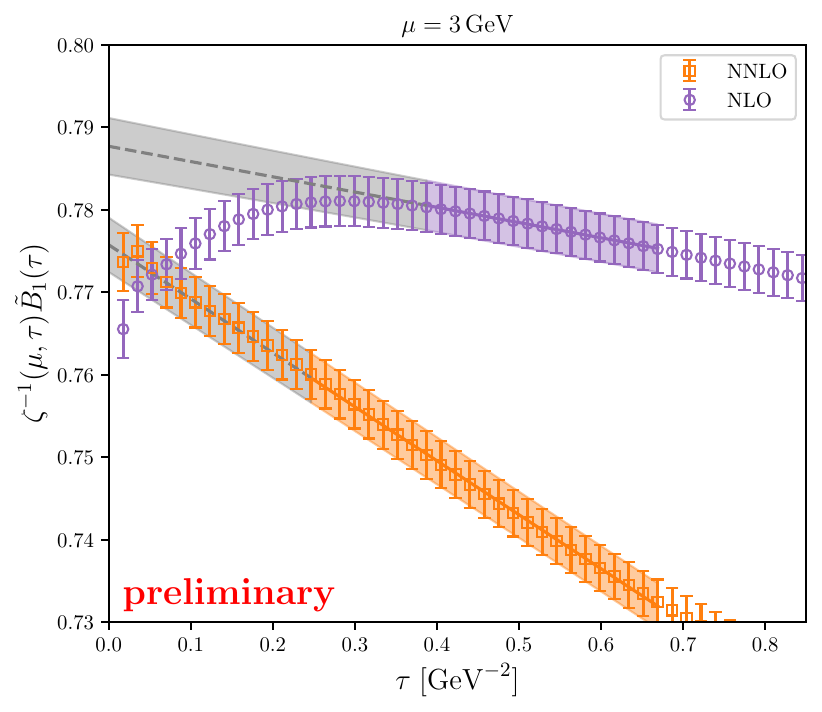}
    \caption{\label{fig:cont_comp} Results for the combination $\zeta^{-1}_{B_1}(\mu,\tau)B_1^{\rm GF}(\tau)$ as a function of flow time, taking NLO (purple) and NNLO (orange) perturbative matching with $\mu=3\,$GeV. Our error bars reflect only statistical uncertainties but we know that for small flow times ($\tau\lesssim 0.2\,{\rm GeV}^{-2}$) the continuum extrapolation contributes a large systematic uncertainty. Using the purple/orange shaded region in flow time, we perform an uncorrelated fit for the $\tau\to0$ extrapolations shown as gray bands.}
\end{figure}

When combining with the data from the lattice simulations, we expect to obtain the $\MSb$-renormalised result by taking the $\tau\to 0$ limit assuming a linear dependence on the GF time. The outcome is shown in Figure~\ref{fig:cont_comp}. The purple circles utilize the NLO matching to $\MSb$-scheme, whereas the orange squares use NNLO matching coefficients. In both cases an extended linear region is present. Using the NNLO coefficients this linear region extends to lower flow times compared to the NLO results. The next step is to seek a window in flow time where the gradient flow has had sufficient effect on the lattice results such that they are renormalised but the flow time is still small enough for higher-dimensional operators to be suppressed. 

If we use as a first guess a flow time window $0.25\,{\rm GeV}^{-2}\leq\tau\leq0.67\,{\rm GeV}^{-2}$ for NNLO and $0.39\,{\rm GeV}^{-2}\leq\tau\leq0.67\,{\rm GeV}^{-2}$ for NLO, we can perform the $\tau\to 0$ extrapolation shown by the gray bands to obtain the renormalised bag parameter in the $\MSb$-scheme at zero flow time. While the difference in the prediction may indicate systematics due to the order of the perturbative matching, we also point out that other systematic effects, e.g.~from the continuum extrapolation, need to be accounted for. In addition we only consider a naive error estimate for the $\tau\to 0$ limit which warrants improvement.
 
Similar discussions regarding the $\tau\to0$ extrapolation for the energy-momentum tensor in the \sftx~framework are given in e.g.~Refs.~\cite{Taniguchi:2020mgg,Suzuki:2021tlr}. 
Further study is still required to fully understand the validity range of both extrapolations.

Phenomenologically, `neutral $D_s$' mixing as is calculated here does not exist, however the results should be similar in magnitude to that of short-distance $D^0$ mixing since any spectator effects are expected to be small. 
In the literature, the short-distance matrix elements for $D^0$ mixing have been calculated on the lattice by FNAL/MILC at $N_f=2+1$ and ETMC at $N_f=2+1+1$, with $\mu=3\,$GeV.
ETMC finds a value of $B_1^{\MSb}=0.757(27)$ \cite{Carrasco:2015pra} (they also have a calculation at $N_f=2$ \cite{Carrasco:2014uya}).
FNAL/MILC quotes a value for $\langle{\cal O}_1\rangle^{\MSb}$; using PDG 2023 \cite{ParticleDataGroup:2022pth}~and Eq.~\eqref{eq:B1}, this leads to $B_1^{\MSb}=0.795(56)$ \cite{Bazavov:2017weg}. 
In Ref.~\cite{Kirk:2017juj}, there also exists a QCD sum rules calculation which, using PDG 2023 \cite{ParticleDataGroup:2022pth}, results in $B_1^{\MSb}=0.636^{+0.091}_{-0.079}$.
One can see in Figure~\ref{fig:cont_comp} that our preliminary results extracted here lie between the two literature values from lattice QCD and slightly above that from QCD sum rules.
While further scrutiny is still required, this is a promising sign for our method as a novel renormalisation and matching-to-$\MSb$ procedure.
It motivates further study of $\Delta F=2$ matrix elements in the \sftx~as a test case towards a calculation of the long-sought-after $\Delta B=0$ four-quark matrix elements.
\vspace{-10pt}

\section{Summary}
\vspace{-10pt}
The $\Delta B=0$ four-quark dimension-six matrix elements are important quantities in accurately and precisely predicting the lifetime of a $B$ meson from the heavy quark expansion. 
Lattice QCD calculations of these matrix elements are strongly sought-after, but no full calculation has been performed to date, with part of the difficulty coming from mixing with lower-dimensional operators under standard renormalisation procedures.
Here we have outlined the idea of using the gradient flow and \sftx~as an alternative renormalisation scheme and matching-to-$\MSb$ method to bypass the issue of operator mixing.
First simulations were carried out with the focus on $\Delta F=2$ operators where results can be validated against lattice calculations in the literature.
Removing additional extrapolations, the initial analysis has been performed at the physical $D_s$ scale.
Preliminary results show promise and consistency with literature values of short-distance contributions to $D^0$ mixing. However further scrutiny on estimating systematic uncertainties is warranted and getting deeper insight in how to choose the flow time window is desired.

In future work, we aim to extend simulations to all lattice ensembles listed in Table~\ref{tab:ensembles}, and also to multiple heavy quark masses and replacing the strange quarks with light quarks. 
This will allow extrapolation to physical $B$ and $B_s$ systems where further validation against $\Delta B=2$ calculations may be done and physical results for the ultimate goal of the $\Delta B=0$ matrix elements can be reached.

\acknowledgments 
\noindent
Measurements were performed using \texttt{Grid}~\cite{GRID,Grid16} and \texttt{Hadrons}~\cite{Hadrons22}.
Computations used resources provided by the OMNI cluster at the University of Siegen and the HAWK cluster at the High-Performance Computing Center Stuttgart. 
This work was partially supported by DeiC National HPC (g.a.  DEIC-SDU-L5-13).
We used gauge field configurations generated on the DiRAC Blue Gene~Q system at the University of Edinburgh, part of the DiRAC Facility, funded by BIS National E-infrastructure grant ST/K000411/1 and STFC grants ST/H008845/1, ST/K005804/1 and ST/K005790/1.  
{M.B., R.H., F.L., O.W. received support from the} Deutsche Forschungsgemeinschaft (DFG, German Research Foundation) through grant 396021762 - TRR 257 “Particle Physics Phenomenology after the Higgs Discovery”.
The work of F.L.~was supported by the Swiss National Science Foundation (SNSF) under contract \href{https://data.snf.ch/grants/grant/211209}{TMSGI2\_211209}.
We thank ECT* for support at the Workshop ``The Gradient Flow in QCD and other strongly coupled field theories'' during which this work has developed.
Special thanks is given to Felix Erben, Ryan Hill, and J.~Tobias Tsang for assistance in setting up the simulation code.

\bibliographystyle{JHEP-jmf-arxiv}
\setlength{\bibsep}{2pt plus 0.3ex}
\bibliography{B_meson}

\end{document}